\documentclass[conference]{IEEEtran}
\newcommand{\mytitle}{Mutation Analysis: Answering the Fuzzing Challenge}

\usepackage{textcomp}
\usepackage{tikz}
\usetikzlibrary{chains,shadows.blur,math}

\usepackage{tikz}
\usepackage{calc}
\def\checkmark{\tikz\fill[scale=0.4](0,.35) -- (.25,0) -- (1,.7) -- (.25,.15) -- cycle;}
\def\bluecheckmark{\tikz\fill[blue,scale=0.4](0,.35) -- (.25,0) -- (1,.7) -- (.25,.15) -- cycle;}

\def\greencheckmark{\tikz\fill[olive,scale=0.4](0,.35) -- (.25,0) -- (1,.7) -- (.25,.15) -- cycle;}
\def\yellowcheckmark{\tikz\fill[purple,scale=0.4](0,.35) -- (.25,0) -- (1,.7) -- (.25,.15) -- cycle;}

\usepackage{microtype}
\usepackage{comment}
\usepackage{amsmath,amssymb,amsfonts}
\usepackage{footmisc}

\usepackage{framed}
\colorlet{shadecolor}{gray!40}

\usepackage{algpseudocode}
\usepackage{algorithm}
\usepackage{algorithmicx}
\usepackage{graphicx}
\usepackage{textcomp}
\usepackage{xcolor}
\usepackage{xspace}
\usepackage{boxedminipage}
\usepackage{float}
\usepackage{soul}
\usepackage{alltt}
\usepackage{caption}

\usepackage{mdwtab}
\usepackage{syntax}
\usepackage[procnames]{listings}
\usepackage{lstautogobble}
\usepackage{qtree}

\usepackage{fancyvrb}

\usepackage{hyperref}
\usepackage{cleveref}

\definecolor{rltred}{rgb}{0.5,0,0}
\definecolor{rltgreen}{rgb}{0,0.5,0}
\definecolor{rltblue}{rgb}{0,0,0.5}
\hypersetup{
colorlinks=true,
urlcolor=rltblue, 
linkcolor=rltblue,
citecolor=rltblue,
}

\Crefname{figure}{Fig.}{Figs.}
\crefname{section}{Section}{Sections}
\crefname{subsection}{Section}{Sections}
\Crefname{Algorithm}{Alg.}{Algs.}

\usepackage{caption}
\usepackage{subcaption}

\usepackage{balance}
\usepackage[inline]{enumitem}
\usepackage{tikz}
\usetikzlibrary{shapes.misc, positioning}

\usepackage{booktabs}

\usepackage{float}
\usepackage{listings}
\floatstyle{plain}

\newfloat{lstfloat}{htbp}{lop}
\floatname{lstfloat}{Listing}

\usepackage{color}

\newcommand{\fuzzer}{fuzzer\xspace}
\newcommand{\MuA}{Mutation Analysis\xspace}
\newcommand{\Mua}{Mutation analysis\xspace}
\newcommand{\mua}{mutation analysis\xspace}
\newcommand{\hiddenbox}{hidden-box\xspace}

\newcommand{\whitebox}{white-box\xspace}
\newcommand{\greybox}{grey-box\xspace}
\newcommand{\Greybox}{Grey-box\xspace}
\newcommand{\Blackbox}{Black-box\xspace}
\newcommand{\Hiddenbox}{Hidden-box\xspace}

\newcommand{\Whitebox}{White-box\xspace}

\usepackage{hyperref}
\hypersetup{
    pdftitle={\mytitle},    
    pdfauthor={et al.},     
    colorlinks=true,       
    linkcolor=blue,          
    citecolor=blue,        
    filecolor=cyan,         
    urlcolor=blue        
}

\definecolor{eclipseBlue}{RGB}{42,0.0,255}
\definecolor{eclipseGreen}{RGB}{63,127,95}
\definecolor{eclipsePurple}{RGB}{127,0,85}

\lstdefinestyle{Python}
{
    basicstyle=\footnotesize\ttfamily,
    numberblanklines=false,
    language=python,
    tabsize=2,
    commentstyle=\color{gray},
    keywordstyle=\bfseries\color{eclipsePurple},
    morekeywords={assert},            
    stringstyle=\color{eclipseBlue},
    procnamestyle=\bfseries\color{black},
    procnamekeys={def},
    columns=flexible,
    identifierstyle=
}

\definecolor{codegreen}{rgb}{0,0.6,0}
\definecolor{codegray}{rgb}{0.5,0.5,0.5}
\definecolor{codepurple}{rgb}{0.58,0,0.82}
\definecolor{backcolour}{rgb}{0.95,0.95,0.92}

\lstdefinestyle{mystyle}{
    fancyvrb=true,
    basicstyle=\footnotesize\ttfamily,
    commentstyle=\color{codegray},
    keywordstyle=\color{eclipsePurple},
    escapeinside={(*}{*)},          
    numberstyle=\tiny\color{codegray},
    stringstyle=\color{codepurple},
    breakatwhitespace=false,
    breaklines=true,
    captionpos=b,
    keepspaces=true,
    numbers=right,
    numbersep=5pt,
    procnamestyle=\bfseries\color{black},
    procnamekeys={def},
    morekeywords={where,assert},            
    showspaces=false,
    showstringspaces=false,
    showtabs=false,
    tabsize=2,
    morestring=[b]' 
}
\usepackage[most]{tcolorbox}
\tcbset{
    left=0pt,
    right=0pt,
    top=0pt,
    bottom=0pt,
    width=4pt+\columnwidth,
    enlarge left by=-2mm,
    enlarge right by=2mm,
    boxsep=5pt,
    }
\tcbset{
    noparskip,
    colback=blue!10,
    colframe=blue,
    coltext=black,
    coltitle=white,
    boxrule=0.3mm,
    fonttitle=bfseries,
}

\usepackage{cleveref}
\def\|#1|{\textit{#1}}
\def\<#1>{\texttt{#1}}

\setenumerate[1]{label={(\arabic*)}} 
\usepackage{xspace}
\newcounter{todocounter}
\newcommand{\todo}[1]{\marginpar{$|$}\textcolor{red}{\stepcounter{todocounter}\footnote[\thetodocounter]{\textcolor{red}{\textbf{TODO }}\textit{#1}}}}

\newcommand{\rem}[1]{\textcolor{red}{\textbf{REMOVED }\st{#1}}}
\newcommand{\done}[1]{\marginpar{$*$}\textcolor{green}{\stepcounter{todocounter}\footnote[\thetodocounter]{\textcolor{black}{\textbf{DONE }}\textit{#1}}}}

\renewcommand{\todo}[1]{}
\renewcommand{\done}[1]{}
\renewcommand{\rem}[1]{}

\ifCLASSINFOpdf
\else
\fi

\begin{document}
\fvset{numbers=left,numbersep=3pt,fontsize=\small,fontfamily=helvetica,frame=lines,resetmargins=true}
%
\title{\mytitle}

\author{
\IEEEauthorblockN{Rahul Gopinath}
\IEEEauthorblockA{CISPA Helmholtz Center\\
for Information Security\\
Email: rahul.gopinath@cispa.de}
\and
\IEEEauthorblockN{Philipp G\"orz}
\IEEEauthorblockA{CISPA Helmholtz Center\\
for Information Security\\
Email: philipp.goerz@cispa.de}
\and
\IEEEauthorblockN{Alex Groce}
\IEEEauthorblockA{Northern Arizona University\\
Email: agroce@gmail.com}
}


%


\maketitle

\begin{abstract}
Fuzzing is one of the fastest growing fields in software testing.
The idea behind fuzzing is to check the behavior of software against a large
number of randomly generated inputs, trying to cover all interesting parts of the
input space, while observing the tested software for anomalous behaviour.
One of the biggest
challenges facing fuzzer users is how to validate software behavior, and how to
improve the quality of oracles used.

While mutation analysis is the premier technique for evaluating the quality of
software test oracles, mutation score is rarely used as a metric for evaluating
\fuzzer quality.  Unless mutation analysis researchers can solve
multiple
problems that make applying mutation analysis to fuzzing challenging,
mutation analysis may be permanently sidelined in one of the most important areas
of testing and security research.

This paper attempts to understand the main challenges in applying mutation
analysis for evaluating fuzzers, so that researchers can focus on solving these
challenges.
\end{abstract}


%
\IEEEpeerreviewmaketitle

\section{Introduction}
\label{sec:introduction}
In the new millennium, fuzz testing (fuzzing) has rapidly become one of the most popular
techniques used in cybersecurity to test the robustness of programs~\cite{manes2019the}. 
It is used by industry giants such as Google~\cite{babic2019fudge,ispoglou2020fuzzgen},
Microsoft~\cite{godefroid2012sage}, Amazon~\cite{mocanu2021fuzz}, Meta~\cite{metafuzz}
and others. Industry behemoths such as Microsoft now mandate fuzzing of every untrusted
interface of every product~\cite{godefroid2020fuzzing}.
Practitioners have used fuzzing to find  thousands of
vulnerabilities~\cite{godefroid2020fuzzing,ding2021an} and other bugs in various
applications.
Indeed fuzzing has in effect democratized the field of software testing.
Organizations such as small businesses can buy off-the-shelf fuzzers,
or make use of online services~\cite{microsoft2021onefuzz}, to evaluate their
software, and have confidence that their applications are free of easily
exploitable vulnerabilities.

While fuzzing has taken the world by storm, several challenges remain. The
foremost among them is the question of oracles~\cite{boehme2021fuzzing}.
Traditional first generation fuzzers typically focused on finding inputs that
made the program under evaluation crash. However, this is no longer sufficient.
As Firefox fuzzing directions~\cite{firefox2022fuzzing} note:
\emph{Fuzzing is only effective if you are able to know when a problem has been found.}
That is, we need fuzzers to be able
to identify more types of vulnerabilities. We need to find bugs that do not only
manifest as a program crash. Going forward, we need to evaluate fuzzers
not just on their ability to cover the program code, but also on their
intelligence in detecting different kinds of incorrect behaviors.

Fuzzing research typically uses two metrics for
evaluating \fuzzer effectiveness: (1) various forms of coverage, with paths taken
seen as the best measure~\cite{bohme2020fuzzing} and branches taken
being another common measure; and (2) artificial benchmarks with seeded bugs
(Magma~\cite{hazimeh2020magma}, LAVA-M~\cite{gavitt2016lava}, BugBench~\cite{lu2005bugbench}, CGC~\cite{lee2015darpa}, Google FTS~\cite{google2020fts}, FuzzBench~\cite{metzman2021fuzzbench}, UNIFUZZ~\cite{li2021unifuzz}, Apocalypse~\cite{roy2018bug}, EvilCoder~\cite{pewny2016evilcoder})
which are often considered the \emph{ground truth}.

However, we know that coverage can often be quickly saturated~\cite{chen2020revisiting},
and that adequate coverage is a necessary, but not sufficient condition for
all bugs to be detected. Indeed, fuzzing researchers are aware of the limitations
of coverage as a proxy measure~\cite{klees2018evaluating}.

Regarding seeded bugs, the basic problem is that the
distribution of such bugs is subject to human bias. That is, engineers who are
tasked with inserting bugs may often be biased about what kind of bugs are
possible and where they may be present. Hence, the distribution of such seeded
bugs need not follow the actual real-world distribution. If one relies on
automatic tools, the bug distribution may also be biased due to the capability
of the tool in question. Indeed, as Bundt et al.~\cite{bundt2021evaluating} notes,
these are limited by the the reachability of their analysis, limited bug types,
and bug realism.

On the other hand, if one uses harvested bugs from existing
programs~\cite{neuhaus2007predicting}, this too
introduces a bias. The issue is that, the availability of bugs does not mean
that the particular program elements contained more bugs. Rather, it only means
that the bugs in those elements were more easily detected. That is, any tool
that performs well in those benchmarks is likely to be better at finding bugs
that we already know how to detect; such an approach may well penalize tools that are
most suited for finding new \emph{kinds} of bugs.

A related problem is that tools such as AFL are commonly used to find and remove
bugs in programs before they become deployed or published. This means that the
bugs found by such popular tools are likely to be found less and less in the
bugs available for harvesting (e.g. in CVEs).  This, in contrast,
rewards tools that may be less able to detect bugs AFL finds easily,
but are able to find bugs AFL cannot find easily.

A final issue in using such benchmarks is that fuzzers can become overfitted
to finding bugs in such benchmarks~\cite{klees2018evaluating,bundt2021evaluating}.
That is, we run the risk that our fuzzers become more and more efficient in
 exploring the bug-inducing statements in the
benchmark, and more effective in triggering the bugs in the benchmark to the
detriment of their real-world performance.

Various proposals for handling these problems have been posed; for
instance Gavrilov et al.~\cite{qrs2020gavrilov} propose using multiple
versions of a program and detecting differences exposed by fuzzers as
a richer evaluation measure (they also provide a more in-depth
examination of the weakness of the coverage and seeded-bug measures
discussed above). However, such an approach requires the availability
of multiple versions of a program, and is not fundamentally tied to
measuring bug detection (if outputs differ but are not
flagged as faulty, this is seen as a difference in appeal,
regardless of oracle strength\todo{This is a bit unclear -- PG}).

Given the importance of oracles in fuzzing, one might expect mutation analysis to
be actively used for \fuzzer evaluation. Mutation analysis is free of
all the above problems that we identified. For example, mutation analysis
is much harder to saturate than code coverage~\cite{chen2020revisiting}, and is
more robust than various forms of coverage as a proxy for the fault
revealing power of the test suite. \todo{We should go through all mentioned
problems and explain why they are not a problem for mutation testing. I think
we do that anyway but not in a clearly structured way. A reader at this point
would likely have to go back and find all problems again. -- PG}

Similarly, the mutations produced by mutation analysis are based on a simple
fault model that correspond fairly well to
real-world~\cite{gopinath2014mutations} faults.
That is, the simple faults produced by mutation analysis are free of influence
from human biases. Further, given that the mutations produced are not based on
any sort of harvested bugs, these are free of bias due to
availability: \mua will induce bugs that are both easy and hard
for, e.g., AFL to detect.

Unfortunately, out of the numerous fuzzing evaluation research papers available~\cite{klees2018evaluating,li2021unifuzz,bundt2021evaluating,paassen2021my,wang2021industrial,tan2019new,qrs2020gavrilov}
\emph{none} recommends the use of mutation analysis for fuzzing. Indeed, none of the papers
we examined~\cite{manes2019the,li2018fuzzing,liang2018fuzzing,chen2018systematic}
actually used mutation score as a means of evaluation.  To our
knowledge, the only published paper that uses mutation to evaluate fuzzing is a Software Engineering in Practice short paper~\cite{icseseip22} discussing an
effort to improve fuzzing for Bitcoin Core (suggesting that
\emph{practitioners} are interested in the possibility, even if
researchers are not, yet).
This is surprising because mutation analysis can actually answer many of the
challenges posed by fuzzing researchers such as computing residual risk~\cite[C.7]{boehme2021fuzzing},
and producing faults that are similar to real bugs~\cite[C.11, C.12]{boehme2021fuzzing}.
Indeed, the theory of mutation analysis is well researched, and mature.


In this paper, we examine the reasons why mutation analysis has, so far,
evaded the attention of security researchers. We identify a few likely
reasons, propose mitigations,
and identify areas of future research.

\section{Background}
\subsection{Fuzzing}
Fuzzing is simple in concept. Given any application that accepts user specified inputs,
the application is executed with 
inputs, generated by the \fuzzer, that try to exercise
as much of the application behaviour as possible.
The program execution is monitored for crashes or other surprising
behavior that can be detected by the available oracles~\cite{manes2019the}.

Fuzzers, and test generators in general are typically evaluated based on
the (1) speed of their turn around, their (2) speed of generation of inputs, and the
(3) quality of their oracles. We define the following metrics for fuzzer evaluation\footnote{
We note that there is no agreement on what these terms mean in fuzzing research. For
example, B\"ohme et al.~\cite{boehme2021fuzzing} defines efficiency as the rate at
which vulnerabilities are discovered, and effectiveness as the total number of
vulnerabilities that a \fuzzer can discover in the limit (i.e. given infinite time).
However, given that code coverage is used to measure \emph{effectiveness} in a majority of
papers~\cite{klees2018evaluating}, we believe that \emph{effectiveness} as used by the
fuzzing community relates to the quality of inputs rather than the oracle used.
Hence, we use \emph{efficacy} to denote the oracle quality here as it seems unused in
fuzzing literature (and we could not find a corresponding term for oracle quality in
fuzzing research). Finally, the speed at which vulnerabilities can be found seems to
be related to the speed at which one can execute the tests
(assuming a uniform distribution and difficulty of finding them).
Hence, we use \emph{efficiency} to denote the speed of execution.
}.
\begin{description}
  \item[Efficiency of the framework.] The turn-around time for the \fuzzer. It
    is influenced by the speed of generation of inputs for
    execution, whether the \fuzzer can be parallelized, the speed of execution of
    the program for a given input.
  \item[Effectiveness of the generator.] The effectiveness of the tests generated in covering 
    all features of the program, and the effectiveness of inputs for triggering
    failures.
  \item[Efficacy of the oracle.] The capability of the \fuzzer oracle to detect triggered
    changes in behavior, especially failures or vulnerabilities.
\end{description}

\subsubsection{Input Generation Techniques}
Fuzzers rely on being able to test the program under fuzzing a large number of
times. They rely on raw computing power to accomplish this fast (that is, within
a given time budget), and there is a lot of focus on improving the speed of
execution of the input generator~\cite{gopinath2019building,srivastava2021gramatron}
as well as fast execution of programs under test~\cite{zalewski2019american}.
As to the \emph{intelligence} used for generation of inputs,
test generators in general are typically classified based on the assumptions they make about the
program in question. These include:
\begin{description}

  \item[\Whitebox generator.]
    \Whitebox generators assume the availability of source code for both static
    as well as dynamic analysis. These generators are also called clear-box generators or
    structural test generators. The availability of source code has traditionally
    meant that the program input specification is also known.
    Sage~\cite{godefroid2012sage} is a canonical example.

  \item[\Greybox generator.]
    \Greybox generators are a spectrum of generators where deep program
    analysis may be impossible or prohibitive to perform. Instead, these
    generators assume that the program can be instrumented, and the
    feedback can be used to guide the program input generation.
    AFL~\cite{zalewski2019american} is a canonical feedback driven \greybox
    generator. 
    A \greybox or \whitebox generator can be used even if the formal program
    interface is not known. Starting with an initial seed corpus, that may
    just be a single empty input, the generator mutates one input of the corpus,
    such as randomly changing a byte, and observing the resulting feedback.
    Interesting inputs are added to the corpus for
    further mutations, a typical evolutionary 
    algorithm. 
    The efficacy of this approach can be enhanced by a seed corpus that
    excerises larger parts of the target program, providing a larger frontier to
    find interesting inputs. Fuzzers that rely on mutation of
    a seed corpus for exploring the input space are called \emph{mutation
    fuzzers}.  AFL is again a canonical example. Both mutation and generation
    can also be combined as in AFLSmart~\cite{pham2019smart}. 
  \item[\Blackbox generator.] \Blackbox generators are used when the source code of the
    program under test is not available, and it is impossible to instrument
    the program for feedback. These generators typically assume that the
    program input specification is known~\cite{gopinath2019building} which is used for
    intelligent generation of inputs\footnote{
\Blackbox testing is also called \emph{specification-based} testing.
}. These kinds of fuzzers are also called
    \emph{generational fuzzers} or \emph{grammar fuzzers}.
  \item[\Hiddenbox generator.] Traditionally called dumb generators, these are
    generators that make the least amount of assumptions about the program in question.
    They assume \emph{nothing} about what kind of inputs are accepted by the program,
    but rely on the fact that program crashes are undesirable.
    Pure random fuzzers~\cite{miller1990empirical} are the best known, but not the
    only kind of \hiddenbox
    generator. Other examples include Anti-random~\cite{malaiya1995antirandom}
    and failure-feedback~\cite{gopinath2020fuzzing} generators.
\end{description}

\subsubsection{Type of Oracles}
Different fuzzers can also be distinguished by their oracle efficacy, as follows.
\begin{description}
  \item[Explicit oracles.] These are contracts that are explicitly specified by
    the practitioner~\cite{barr2014oracle}. These are also called \emph{specification based}
    oracles~\cite{nardi2015survey}.
    An example of an explicit contract is that the output should conform to,
    say, a given XML schema, or that the output should be a valid JSON file.
    Property based testers such as QuickCheck~\cite{claessen2011quickcheck} 
    can make use of sufficiently detailed contracts for strong oracles.
  \item[Implicit oracles.] Implicit oracles are oracles that describe the
    general valid state of the program. These are the typical oracles used in
   general fuzzers~\cite{barr2014oracle}.  Tools such as
   DeepState~\cite{deepstate} make it possible to use QuickCheck-style
   property-based specifications with off-the-shelf fuzzers like AFL.
   Sanitizers are one of the most common types of \emph{implicit oracles} used for fuzzing.
   A number of sanitisers exist that can
   be used with general fuzzers~\cite{song2019sok}.
   The most popular ones~\cite{jeon2020fuzzan}
   are memory (MSAN) sanitiser which detects uses of uninitialized memory,
   address (ASAN) sanitizer which targets use-after-free buffer-overflows and memory leaks,
   thread sanitizer (TSAN) which detects data races and deadlocks,
   floating-point sanitizer (NSAN),
   and undefined behaviour sanitizer (UB-San)~\cite{song2019sok,courbet2021nsan}.
  \item[Differential oracles.] These are also called
   pseudo oracles~\cite{mckeeman1998differential,weyuker1982testing,davis1981pseudo}.
   If there exists a different program that implements the same specification, then one
   can make use of a differential oracle. The idea is to generate inputs and compare
   the behavior of different programs. If there is a difference between the two implementations,
   at least one of them is wrong.
   Common examples include parsers for common file formats, virtual machines such as CLR, WASM, and JVM,
   and protocol implementations such as FTP and HTTP.
   A common variant of this idea is the regression oracle~\cite{evans2007differential}. In this case,
   a different version of the same program is used to verify that only expected features differ.
   Finally, tools
   such as Evosuite~\cite{fraser2011evosuite}
   that make use of mutants for generating test cases also use the same idea.
  \item[Metamorphic oracles.] These are oracles that rely on metamorphic
  relations between inputs~\cite{chen2016metamorphic,segura2018metamorphic,zhou2004metamorphic,segura2016survey}
  An important category of metamorphic oracles is cross-checking oracles~\cite{carzaniga2014cross},
  which use redundant operations in software APIs as a way of constructing possibly
  equivalent API call test sequences and assertions.
  \item[Dynamic invariants.] The final approach involves monitoring the invariants between program variables
during execution, and tries to identify outlying executions. These are exemplified by Daikon~\cite{ernst2007daikon}.
\end{description}

\subsection{\MuA}
\Mua is a technique for evaluating the fault revealing power of test suites on
a given program.
We define the following mutation related terms:
\begin{description} 
\item[Mutation.] A mutation is a small syntactic change that can be induced in the program, that will likely lead to a semantic difference.
\item[Mutation operator.] Mutation operators are replacement patterns that describe how
mutations are induced in the program. A mutation operator, when applied to a matching location in the program, will produce a \emph{mutant}.
\item[Mutant.] A mutant is a new program that contains differences (mutations) from the
original. A \emph{first order} mutant contains only a single mutation. A \emph{higher order}
mutant contains multiple mutations.
\item[Trivial mutants.] These are mutants that can be killed (detected as failure inducing) without targeted intelligence.
That is, an input that covers the mutation location in the original program already killed them, such as a \fuzzer{}'s seed input.
\item[Redundant mutants.] A mutant A is redundant with respect to another mutant B if any test that can kill
mutant B is guaranteed to kill A.
\item[Duplicate mutants.] A mutant A is duplicate of another mutant B if any test that kills A will kill B and vice versa.
\item[Stubborn Mutants.] These are mutants that remained alive even after coverage reached their mutation locations.
\item[Intelligent mutants.] These are mutants that were killed by fuzzers on individual evaluation (i.e., they required intelligence to kill).
\item[Immortal mutants.] Mutants that can't be killed by any weak oracles
such as crash oracles in the current program.
\item[Equivalent mutants.] An equivalent mutant is a mutant that, while different from
the original program syntactically, has the same semantics.
\end{description}
\subsubsection{Error Model}
For \mua, we start with the following \emph{error model}: Any token\footnote{A token is the smallest syntactical element in the program.} in a program is a possible location for a fault to exist, and faults are likely caused during transcription of the concept in the developer's mind to the code artifact. 
Further, we assume that the developer uses automatic tools such as compilers which can identify and remove a limited category of faults. This gives us a way to generate possible faults without human bias: Simply generate all possible instances of faults for each source code element that will get past the compiler. 
Unfortunately, this can lead to a combinatorial explosion. Hence, we rely on a few axioms
to limit the number of faults generated. The \emph{finite neighborhood hypothesis} and the \emph{coupling effect}.

\subsubsection{Fundamental Axioms}
The \emph{finite neighborhood hypothesis} states that faults, if present in the program, are
within a limited edit distance away from the correct formulation~\cite{gopinath2014mutations}.
The \emph{coupling effect} claims that simple faults are coupled to complex faults, such that tests capable of detecting failures due to
simple faults will, with high probability, detect the failures due to complex faults.
Hence, the probability of fault masking is very low~\cite{offutt1992investigations}.
Both these axioms are well researched,
with well-founded theory~\cite{wah2000theoretical,wah2001theoretical,wah2003analysis,gopinath2017the}, and confirmed in large number of
real world software~\cite{just2014are,gopinath2017the,petrovic2021does,chekam2017empirical}.
With these two axioms, we can limit the faults that we need to test.
Allowing us to focus only on changes to the smallest program elements, such as tokens and statements, and still
expect that the created mutations are representative of real bugs.

Given this error model, the idea of \mua is to simply collect possible fault patterns
(a single fault pattern is called a \emph{mutation operator}). Identify possible faults in the program
(called \emph{mutations}), generate corresponding faulty programs (called \emph{mutants}) each containing a single
\emph{mutation}, and finally evaluate each \emph{mutant} separately using each software verifier (test suites, static analyzers~\cite{qrs2021groce},
static test generators, and fuzzers) and check whether the verifier is able to detect the changed behavior of the mutant
(called \emph{killing the mutant}).  The idea is that the number of mutants thus killed by each verifier provides a simple and effective criteria to compare
verifiers.

\section{Mutation Analysis for Fuzzing -- Advantages}
B\"ohme et al.~\cite{bohme2020fuzzing} identified the following questions as the
current challenges in fuzzing. We mark each question that \mua or \mua research can
potentially answer (\bluecheckmark), help answer (\greencheckmark),
evaluate (\yellowcheckmark), or provide insights on(\checkmark).
\begin{itemize}
  \item[C.1.] How to fuzz more types of software.
  \item[\textbf{C.2.}] How to identify more types of vulnerabilities. \yellowcheckmark
  \item[\textbf{C.3.}] How to find more deep bugs. \checkmark
  \item[\textbf{C.4.}] What kind of vulnerabilities are not found by fuzzing. \yellowcheckmark
  \item[C.5.] How to leverage the auditor.
  \item[C.6.] How to improve the usability of fuzzing tools.
  \item[\textbf{C.7.}] How to assess the residual security risk. \greencheckmark
  \item[C.8.] What are the limitations of fuzzing.
  \item[\textbf{C.9.}] How to evaluate more specialized fuzzers. \bluecheckmark
  \item[\textbf{C.10.}] How to prevent overfitting to a specific benchmark? \greencheckmark
  \item[\textbf{C.11.}] Are synthetic bugs representative? \bluecheckmark
  \item[\textbf{C.12.}] Are bugs discovered by fuzzers, representative? \bluecheckmark
  \item[\textbf{C.13.}] Is coverage a good measure for \fuzzer effectiveness? \bluecheckmark
  \item[\textbf{C.14.}] What is a fair time budget? \yellowcheckmark
  \item[C.15.] How to evaluate techniques instead of implementations.
\end{itemize}

\Mua can help in overcoming 10 out of 15 challenges in fuzzing.
Next we discuss how and why this is the case for each challenge in more detail.

\subsubsection*{\textbf{C.2.} How to identify more types of vulnerabilities.}

This is a major limitation in current fuzzing benchmarks~\cite{bundt2021evaluating}.
As discussed in the introduction, this is the most obvious way
mutation analysis can aid fuzzing research.  Research currently uses
either structural coverage (which almost always provides no insight
into oracles and vulnerability types) or seeded/harvested bugs,
where adding new categories of bug is labor-intensive and prone to
either human or tool bias.

\Mua has a lot to offer here. In particular, \mua is fault-based, which means
that it explicitly evaluates the oracular strength. Something structural coverage metrics
are not able to do, additionally, \mua does not get saturated as fast as coverage does.

Indeed, \mua is one of the best fault based evaluation techniques we have, as it
is based on a well founded theory, and was created with the explicit purpose of
avoiding the pitfalls of fault seeding (i.e. benchmarks).

\begin{tcolorbox}
\Mua avoids pitfalls of fault seeding benchmarks, and can induce a much larger variety
of bugs that match the real world bugs in behavioral variety and difficulty.
\end{tcolorbox}

\Mua exhaustively seeds the program with first order variations. As a consequence,
there is no bias in the kinds of bugs that are seeded,
avoiding the possibility of fuzzers overfitting.
All in all, \mua can adequately answer the challenge posed
by B\"ohme~\cite[C.2]{boehme2021fuzzing}.

\subsubsection*{\textbf{C.3.} How to find more deep bugs.}

Solving this challenge can likely be achieved by improving the quality of input generators. However, inducing
subtle faults into predicates as T-Fuzz~\cite{peng2018tfuzz} does can help. However,
the induced faults need to be small enough so as to affect the validity of the
input execution minimally. Mutation analysis research can help
here~\cite{visser2016what}; in work-in-progress (citation blinded for review),
we have used mutation to generate program variants to explore deep behavior.
Furthermore, unlike traditional fault-injection techniques such as
LAVA-M~\cite{gavitt2016lava}, faults injected by \mua is not limited to a limited
reachable subset of the program as it faithfully recreates possible errors by the
programmer anywhere in the program. Hence, bugs induced by \mua can distinguish
the capability of fuzzers to detect deep bugs.
\begin{tcolorbox}
The faults injected by \mua is not limited by any analysis.
\end{tcolorbox}

\subsubsection*{\textbf{C.4.} What is the nature of vulnerabilities that are not found?}

This challenge is about identifying the nature of bugs that evaded detection
after long fuzzing campaigns. While empirical studies are certainly helpful
here, \mua has something to offer as well. In particular, \mua is a mature field
with well established research on the nature of bugs~\cite{shamshiri2015automatically,just2014are}.

Second, \mua has a well founded theory based on exhaustively seeding first order
faults. The first order faults serve as the base case and the coupling effect
hypothesis which serves as the induction to ensure that a large majority
($>99$\%~\cite{gopinath2017the}) of the higher order faults are found by
test suites adequate to detect these first order mutants. Further, for those
rare faults that slip through, mutation analysis can be easily extended with
new operators that need not be first order~\cite{jia2008constructing,harman2010manifesto}.

\begin{tcolorbox}
The mutants that remain after a fuzzing campaign indicates the kinds of bugs a
fuzzing campaign missed.
\end{tcolorbox}

What about failures that require interaction between multiple faults?
Such faults are impossible to find by first order mutation. However, there are
two mitigations here. The first is that
surpricingly small number of faults are sufficient for a majority of FTFI
(failure-triggering fault interactions) found~\cite{kuhn2004software,gopinath2017the}, with more than 90\% of the
failures observed involving just one or two faults. Indeed, most failures were
triggered by a single erroneous parameter to a function, and almost all could
be induced by fewer than 4 faults.
Second, one may induce subtle mutations using higher order mutation~\cite{jia2008constructing}
which can adequately represent such rare faults.


\subsubsection*{\textbf{C.7.} How to assess the residual security risk?}
One of the main reasons for using \mua is that it provides the best estimate for 
\emph{residual defects} in a program, the number of defects that remain in
a program after testing is completed\footnote{At this point, we assume
that no \emph{detected faults} remain in the program. Any that were found
were fixed.}. Assume for a moment that the fault
was small enough to be a mutation,
and one of our mutants reverses this fault. If so, (assuming sound
testers), a testing professional
would never have written a test to kill the mutant.
Rather, it would have been flagged a defect, and fixed.
If, on the other hand, the
difference between the corrected version and the faulty version was bigger than any mutant,
by the coupling effect hypothesis,
there must still exist multiple mutants that correspond to different parts of the failure
behavior (not necessarily representing the correct behavior).
If so, for a tester to
kill such mutants, the tester has to assert the correct behavior for that mutant,
and given that the failure behavior is different from the correct behavior, this would
again have been flagged as a fault and fixed.
This is true for assertions from any oracle, so long as the oracle is sound.
We note that fuzzing relies on weak crash oracles that are sound but not complete.
Hence, as the number of mutants that remain undetected decreases,
the number of defects that remain undetected also decreases monotonically.
Hence, the number of mutants that remain alive is
a \emph{true ordinal measure}\footnote{
By true ordinal measure, we simply mean a measure that corresponds to the definitions from
measure theory~\cite{tao2011introduction}. That is, it follows monotonicity of the
measured quantity, and additivity of measures between independent subsets. That is,
it is different from mere correlation.}
for the residual defects in the program.

\begin{tcolorbox}
\Mua can comprehensively answer the question of residual risk after fuzzing.
\end{tcolorbox}

\subsubsection*{\textbf{C.9.} How to evaluate more specialized fuzzers.}
This question is about the challenge of evaluating fuzzers that are focused on
specific kinds of bugs, or specific kinds of programs. For example, one may be
out of luck if one is looking for parser bugs, or specifically binary parser
bugs because the baselines may not contain such parsers.
As B\"ohme notes~\cite{boehme2021fuzzing},
current benchmarks are often not designed for such tasks, and he advocates for
suitable programs and baselines for comparison. The trouble is that, adding new
programs to your benchmark is only postponing the problem. Say you have --
after many months of manual effort -- added a few parsers and corresponding
harvested bugs to your benchmark.
A new language feature, or a new kind of parsing such as combinatorial parsing
can make such benchmarks inadequate to deal with the new kinds of bugs
introduced, while removing the bugs that are evaluated in your benchmark. Hence,
improving the benchmark is not a solution here.

However, \mua has a simple solution here. If a particular kind of program is
not present in the benchmark, simply add the program, and let \mua generate the
bugs to be evaluated. This removes the manual
effort present in keeping the benchmark up to date.

\subsubsection*{\textbf{C.10.} How to prevent overfitting to a specific benchmark?}
In this question, B\"ohme points out that fuzzers can become overfitted to a
specific benchmark irrespective of the superiority of a benchmark. B\"ohme
proposes various solutions involving collecting even more benchmarks. The
trouble here is again that for fault-seeding, one needs not only programs, but
also harvested or seeded bugs on these programs. This is a labour intensive
task.

As before, the \mua solution is simple. Collect as many programs as are
possible. So long as you are able to run it, \mua will take care of seeding
the right bugs for you. Indeed, if you are a user of specific kinds of programs
such as the military or government, with the traditional approach of fault seeding,
it is nearly impossible to find preexisting benchmarks that suit your purpose.
With \mua, you can make your own benchmark, and evaluate fuzzers on it without
much manual effort.

Further, one can extend \mua with new kinds of mutation operators as well as higher
order mutants~\cite{jia2008constructing,harman2010manifesto} deriving more value out of an existing benchmark.

Similarly, if you are working on a new or less popular language, you are again
out of luck with the traditional fault-seeding benchmarks.  With \mua, you can simply collect
programs, and \mua to seed bugs, given the existence of
\emph{universal} mutation tools that apply to essentially any program language~\cite{groce2018extensible}.

\begin{tcolorbox}
Using \mua fuzzing researchers can avoid overfitting to benchmarks.
\end{tcolorbox}

\subsubsection*{\textbf{C.11.} Are synthetic bugs representative?}
This question is about whether the seeded bugs designed by engineers
or by artificial injectors such as LAVA-M~\cite{gavitt2016lava} are
representative of the real bugs~\cite{bundt2021evaluating}.
If not, what can we do to make them so?
This is a question that was comprehensively answered by \mua research.
Indeed, we know that the faults seeded by \mua are representative of real world
bugs both in terms of syntax~\cite{gopinath2014mutations} as well as
semantics~\cite{andrews2005mutation,andrews2006using}, and having a higher mutation
score means lesser number of found bugs in the future~\cite{ahmed2016can,papadakis2018are}, and
thus mutants are valid substitutes for real faults~\cite{just2014are}.
That is, if one is to rely on bugs induced by \mua, one can have high confidence
that the bugs induced are representative of the real world.

\begin{tcolorbox}
The bugs induced by \mua can match the realism of real world bugs.
\end{tcolorbox}

\subsubsection*{\textbf{C.12.} Are bugs discovered previously by fuzzers, representative?}
This question is about bugs collected through harvesting preexisting bugs for
benchmarks. As we discussed in \emph{C.11}, the faults induced by \mua are
indeed representative of real world. Further, new fault patterns can be easily
mined and added if necessary.

\subsubsection*{\textbf{C.13.} Is coverage a good measure for \fuzzer effectiveness?}

The problem with coverage is that
it is a necessary but not sufficient condition for detecting faults. In
particular, covering a fault is a necessary but not a sufficient precondition
for triggering that failure, and even if the failure is triggered, we need
adequate oracles that can detect the failure. Coverage does not measure the
quality of oracles available. Secondly, coverage is often easily saturated~\cite{chen2020revisiting}. Once coverage is saturated, there is little
feedback available for further fuzzing.

\subsubsection*{\textbf{C.14.} What is a fair time budget?}
In this question B\"ohme points out that there may be a difference between
fuzzers in terms of the number of vulnerabilities that they can detect in
the limit (given infinite time).
However, even a \hiddenbox \fuzzer assuming nothing about the program
in question can, in the limit, generate any input. Any \fuzzer that doesn't can
be trivially extended with a \hiddenbox \fuzzer to be maximally effective
in terms of its inputs. However, the idea here seems to have been about
distinguishing between fuzzers in terms of the trade-off between efficiency
and effectiveness in terms of the curve of discovery.
We note that the complexity of a program is related to the number of mutants
it can have~\cite{parsai2020mutant}. Hence, any fair budget should correspond
to the number of mutants produced.
We discuss how to measure this curve in more 
detail in \Cref{sec:distribution}.

\tcbset{
    noparskip,
    colback=green!10,
    colframe=green,
    coltext=black,
    coltitle=white,
    boxrule=0.3mm,
    fonttitle=bfseries,
}

\begin{tcolorbox}
\Mua can adequately answer or help answer, evaluate, or provide insights on 10 of the 15 major challenges faced by fuzzing research.
\end{tcolorbox}
\tcbset{
    noparskip,
    colback=red!10,
    colframe=red,
    coltext=black,
    coltitle=white,
    boxrule=0.3mm,
    fonttitle=bfseries,
}

\section{Mutation Analysis for Fuzzing -- Challenges}

Mutation score and structural coverage metrics are typically used for two
different but related purposes in the software-engineering world.
\begin{enumerate}
  \item The first, and
traditional, use is by a practitioner to evaluate the \emph{adequacy} of a test
suite. That is, if one considers statement coverage, a coverage result of 100\%
means that all statements in the program were executed at least once. That is,
the test suite is coverage adequate.
\item The second common use of mutation score and other coverage metrics is as a
means of \emph{comparing between two test suites}.
\end{enumerate}
The \emph{adequacy} of a test suite was often the most important factor when
considering static test suites. The reason is that, in traditional testing
scenario, one rarely needed to compare different ways of creating test cases,
because there was often only one kind -- developer written test cases. However,
with the advent of test generators such as Evosuite~\cite{fraser2011evosuite},
Randoop~\cite{pacheco2007randoop}, and different fuzzers, the situation has
changed. Most of these fuzzers and other test generators can make use of any
and all computational resources as you can provide them. There is no fixed point
at which they stop. Unfortunately, there is an exponential increase in difficulty in finding
the next bug~\cite{bohme2020fuzzing}. Hence, we need to decide how to allocate
limited testing budget, and for that we need to find the best test generator.
\subsection{Computational Expense}
\label{subsubsec:computation}
The main problem with \mua is its computational requirements. The number of mutants that have to be
evaluated increases with the size of the program. Further, for fuzzing we need to evaluate each input
produced independently on each mutant. That is, we can't tell if a mutant will be killed by an
input without executing the mutant on that input. We can't even assume that the fuzzer will produce the same input on the
original as well as the mutant because the fuzzer may find the mutation in the program using
static analysis, and take steps to reach it or to induce failure on a perceived fault.
Further, most feedback driven fuzzers (e.g.,  AFL) will modify their next input correspondingly to
take advantage of any new coverage found, which may be induced by the mutation. Hence, it is highly
likely that the fuzzer generated inputs are different for the original
code and various mutants.
This means that there is a quadratic increase in the number of program executions required with program size.

There are a number of traditional optimizations to make mutation
analysis less computationally demanding. 
However, most of these techniques assume static test suites, 
which makes them inapplicable to fuzzers.
For example, two algorithms try to reduce the number of mutants to be evaluated by
identifying which tests are applicable to which mutants. The first, called lazy mutation
analysis, by Fleyshgakker~\cite{fleyshgakker1994efficient}
uses weak mutation kills to identify which tests to run on which
mutants. A less complex approach is to use code coverage for the same purpose~\cite{schuler2009javalanche,mateo2015reducing}.
The idea is to find the statements in the program that are covered by the specific
tests in the test suite, then only run tests against mutants they cover.
Unfortunately, these techniques are inapplicable
to fuzzers because fuzzers may not use the same input sequences on all
mutants, as we
mentioned above.

That is, a fuzzer may analyze the source code statically and modify its
behavior correspondingly resulting in different fuzzer behaviors for the original program and the
mutant (violating the clean program assumption ~\cite{chekam2017empirical}).
Another optimization is to use \emph{weak mutations}~\cite{howden1982weak} which only check whether
an input would have resulted in a different state in the mutant when compared to the original. This
again assumes that the test suite will not change when a mutation is present.
Partitioning infected states~\cite{just2014efficient},
split stream execution~\cite{tokumoto2016muvm, gopinath2016topsy}
and equivalence modulo states~\cite{wang2017faster} assume again that
the same
inputs are used for both mutant and original program.
A recent optimization technique is to memorize expensive methods~\cite{ghanbari2021toward} so that
if function calls with the same arguments are present in the original and the mutant, then the results
can be reused. The effectiveness of this technique again depends on the same inputs being used for
original and mutant.
Indeed, even the relatively harmless schemata based mutation execution
 optimization~\cite{untch1993mutation,mateo2012mutant,just2011using} can
induce changes in the fuzzer behavior, as the fuzzer can wastefully try to generate inputs
in an attempt to reach the disabled mutations when another mutation is enabled.

Techniques such as test selection~\cite{chen2018speeding,zhang2013faster} can't
be used for fuzzing because they rely on a static test suite.

Thus, these traditional optimization techniques do not work well for fuzzers.

\begin{tcolorbox}
  \textbf{Challenge 1}: Find optimizations that do not rely on static test suites.
\end{tcolorbox}

\subsection{Comparing Test Generators}
\subsubsection{Non-determinism}
\label{sec:nondet}
One of the challenges in evaluating test generators is that many of these are
dependent on non-deterministic execution for their effectiveness. The problem
with non-determinism is also that the results achieved depend on the initial
random seed, and also on the seed corpus~\cite{klees2018evaluating}.
Hence, one evaluation alone is insufficient for statistical confidence in the
evaluation rankings~\cite{klees2018evaluating}.

Unlike coverage techniques which impose very little overhead over the program,
\mua can be costly.
Hence, running \mua multiple times for statistical certainty can be prohibitive.

\subsubsection{Distribution}
\label{sec:distribution}
The second challenge in evaluating test generators is that there is no single
point at which a test generator stops. Secondly, the speed of analysis and
execution plays a large role in how effective a \fuzzer is. For example, a
naive \fuzzer~\cite{miller1990empirical} or a simple grammar
\fuzzer~\cite{gopinath2019building} can run circles around a more intelligent
symbolic execution based \fuzzer in the initial stages~\cite{godefroid2012sage}
because program analysis and symbolic execution can take time. However, beyond
the easily explorable program paths, the symbolic \fuzzer can leverage
information about the predicates guarding program paths, and make better use
of the computational resources at a later stage~\cite{klees2018evaluating}.
Finally, any \fuzzer can (or can be trivially extended with a
\hiddenbox \fuzzer to)
find all bugs that its oracle can detect, in the limit. This is because, in the limit,
all inputs will be tried at some point.

Hence, we need to identify the fault revealing curve of a test generator for
adequate comparison, which should be provided to the practitioners.
A single score will no longer cut it.

However, unlike coverage which can be obtained from a single program, multiple
mutants need to be evaluated for \mua. This means that there can be impacts
due to parallelization, scheduling algorithms etc., on the average number of
mutants detected at any point in time.







\begin{tcolorbox}
  \textbf{Challenge 2}: Find ways to incorporate statistical and time based
  distribution of mutant kills cost-effectively.
\end{tcolorbox}

\subsection{Problems with Redundancy}
One of the problems with \mua is that the mutants produced may be redundant
or there may even be true duplicate mutants. The problem with such mutants is that
they can affect the correlation of mutation score to the real fault detection.
That is, say you have N duplicate mutants. If we have an input that kills these
N mutants, it is only worth one mutant, but will be counted as N in the mutation
score. On the other hand, if these mutants remain undetected, the number of
 mutants yet to be killed is inflated. Hence, redundant and duplicate mutants are
undesirable. The traditional approach to eliminating them has been to compute the
full mutation matrix of tests x mutants result, and compute the minimal mutant
set. However, with fuzzing this is not possible for two reasons: (1) there is
no end point for fuzzing. Hence, the size of test suite is limited only by the
available time and (2) the test suites may differ between each mutant. That is, we
have no simple way to compute duplicate and redundant mutants during fuzzing.

\begin{tcolorbox}
  \textbf{Challenge 3}: Determine how to account for redundant mutants during fuzzer comparison.
\end{tcolorbox}

\subsection{Equivalent Mutants}
A somewhat similar problems with \mua is equivalent mutants. These are mutants that
are semantically the same as the original program. For example, given a cache optimization
for a complex computation such as below,
\begin{lstlisting}
if (cache.has(key))
   return cache.get(key);
return compute(key);
\end{lstlisting}
removing the cache check need not
induce a failure.
The problem with equivalent mutants is that they make the final mutation score unreliable.
That is, we do not know for sure if the live mutants remaining are actually killable or not. Indeed,
anywhere from 10\% to 40\% of generated mutants could be equivalent~\cite{jia2010analysis,grun2009impact}\footnote{Later analysis suggests a more refined figure of 23\%~\cite{yao2014study}.}. However, the actual percentages are very program dependent.

One of the promising approaches toward estimation of equivalent mutants is
using \emph{species richness estimation}  as proposed by B\"ohme.~\cite{bohme2019assurances}.
The idea is to use the frequency of counts of mutants that are found by different
test cases to estimate the number of mutants that are yet to be found. This
can provide us with an estimate of the total \emph{killable} mutants if one is
given the full kill matrix that describes which mutants are killed by which tests
(M $\times$ T). Unfortunately, given that the test cases are not static between
different mutants, this method can't be used for estimation of equivalent
mutants during fuzzing.

While equivalent mutants are troublesome for interpretation of mutation score,
such equivalent mutants are less of a concern for fuzzer comparison, because,
for comparison, the only concern is the number of relative mutant kills between fuzzers.
\todo{PG: We propose a solution for this challenge in section V, mention that here?}
\done{
It does have an effect as if one fuzzer finds a single mutant and another finds a group of equivalent mutants
that the respective other fuzzer does not find. The one finding the equivalent mutants will be rated higher. -- PG
Equivalent mutants can't be found!. -- RG
Also if it's not a concern why do we pose it as a challenge? -- PG
Challenge only in only in fuzzer evaluation. But I agree, less of a concern is a better phrasing. -- RG}


%

\begin{tcolorbox}
  \textbf{Challenge 4}: Determine how to account for equivalent
  mutants in fuzzer evaluation.
\end{tcolorbox}

\subsection{Lack of \mua Frameworks that Focus on Fuzzers}
One of the difficulties encountered when trying to use mutation analysis with
fuzzers is that there are no mutation frameworks that are capable of running
fuzzers on mutants efficiently. Fuzzers typically rely on framework support
for coverage feedback. Further,  many fuzzers also rely on starting seeds.
Accounting for, and managing such seeds while also parallelizing mutation
analysis is not yet available in mutation frameworks.

\begin{tcolorbox}
  \textbf{Challenge 5}: Build \mua frameworks that are fully capable of fuzzer evaluation.
\end{tcolorbox}

\subsection{Lack of Awareness Among Researchers}
While we have identified possible technical reasons for the dismal popularity
of mutation analysis among security researchers, lack of awareness may also be
an equally important reason. Indeed, none of the fuzzing review papers we
examined contained any reference to mutation analysis.


\begin{tcolorbox}
  \textbf{Challenge 6}: Improve visibility of \mua among fuzzing researchers.
\end{tcolorbox}

\section{Promising Directions for Research}
\subsection{Computational Expense}
Computational expense in running \mua is one of the major impediments for
the use of \mua to evaluate fuzzers. We discuss a few possibilities to
mitigate this problem.

\subsubsection{Mutant reduction}
One of the possible ways to reduce the cost of \mua without assuming
anything about the test suite is to reduce the number of mutants
evaluated. Indeed, this is one of the traditional techniques for reducing
the cost of \mua. However, one needs to be careful how the number of
mutants is reduced. In particular, mutation operator selection techniques
can have unexpected disadvantages~\cite{gopinath2017mutation}. The best
reduction can be achieved using random sampling or various forms of strata
sampling techniques with different strata such as program elements or
operators.

Sampling may not sound promising because it comes with associated
non-determinism and loss of distinguishing power for \mua (when we sample, we
are essentially accepting a more limited accuracy for the metric).

\subsubsection{Splitting the evaluation}
\label{sec:splitting}
However, there may be a better way out. It may be observed that during
fuzzing, one needs both effectiveness as well as efficacy. That is, one
needs to generate inputs that can cover all possible input features. Further,
we need oracles that can validate these inputs. Given that coverage is good
in the first part, and \mua is good at the second part, why not split them?

The idea is to use coverage as the evaluator for fuzzers until coverage is
saturated. We use the standard 24 hours timeout for fuzzing~\cite{klees2018evaluating}.
Once coverage is saturated for all fuzzers, we collect any and all inputs that
are required to cover the program maximally, and minimize them to produce a
minimal test suite with the same coverage. Let us call this the \emph{coverage seed}.
Next, we use this test suite as a static test suite, and run \mua with it,
removing any mutant that is killed
using this set of inputs. At this point, all traditional optimizations of \mua
can be applied. This allows us to remove \emph{trivial mutants} cheaply.
We then run \mua for each fuzzer starting with the minimal test suite as the
seed corpus for each. With this technique, we only evaluate
\emph{intelligently selected
mutants} with fuzzing runs.

\subsubsection{Supermutants}
A third way out is to evaluate multiple mutants at once. The idea is to produce
higher order mutants where the individual mutations are independent of
each other in terms of semantics~\cite{gopinath2018if} and fuzz the higher order
mutant. The idea is to produce higher order mutants such that any input will cover
at most one simple mutation in the higher order mutant.
Count any crash as killing the mutant corresponding to the mutation that was
covered by the crash inducing input.
The higher order mutant is completely killed when all corresponding simple mutants are killed.
This can reduce the number of mutants to be independently evaluated.

\subsection{Comparing Test Generators}
For comparing test generators, we need the curve of discovery as 
we discussed in \Cref{sec:distribution}. However, if we split the evaluation
into a coverage part and a \mua part as discussed in \Cref{sec:splitting}, then
we can reduce the computational requirements for  computing the curve as follows.
For every \emph{chosen mutant}, we keep track of the time at which it
was killed. We also have the coverage curve which is comparatively easy to obtain,
and the corresponding lines of code that were covered.

Now, for computing the mutation curve, the idea is as follows. For any trivial
mutant, the time taken for detection is the time taken for covering the corresponding
program element. For any chosen mutant, the time taken for killing is the time
taken for covering its corresponding mutation along with the time taken to kill
it. Once we compute the time to kill for each mutant, it can be plotted, and the
curve of mutant detection can be obtained for any time period.

\subsection{Redundancy}
For evaluating the redundancy of mutants, we propose the following solution.
For any mutant, identify the input that kills it during fuzzing. This may be
from the \emph{coverage seed} collection or during later fuzzing. Next, use
this final set of inputs as a static test suite
(let us call this the \emph{final static test suite}),
and run full matrix \mua using the traditional optimizations.
This will allow us to compute
the minimal set of mutants using the traditional minimal mutant computation.

\subsection{Equivalent Mutants}
Once we have the final static test suite, we propose to use the
\emph{final static test suite} along with 
\emph{species richness estimation} by B\"ohme~\cite{bohme2019assurances} for
evaluating the range of equivalent mutants. However, we note that this is yet
to be validated even for traditional \mua.  A second possibility is to
sample from the remaining mutants to evaluate whether the sampled mutant is
killable.
We note that detecting equivalent mutants will be no different for fuzzers
than in traditional \mua.



\section{Related Work}

The latest research in \mua is discussed by Papadakis et al.~\cite{papadakis2019mutation}
who also discuss numerous techniques to reduce the computational expenditure involved in
traditional \mua.

The survey paper by Pizzoleto et al.~\cite{pizzoleto2019systematic} focuses on
cost reduction of \mua and suggests that one of the main areas of research in this
direction recently has been in trying to reduce the
number of mutants executed. Other options include various forms of selective mutation,
statistical sampling of mutants, clustering and then sampling mutants, finding
subsuming mutants, and so forth.
Researchers have also explored various strategies for cost reduction including higher order mutation,
weak and firm mutation~\cite{offutt1993experimental,howden1982weak,offutt1991strong,offutt1994empirical,durelli2012toward}. 
Finally, another possibility is to use a proxy for mutation score such as
checked coverage~\cite{schuler2013checked}.

Lima et al.~\cite{lima2016evaluating} examines different strategies including
different higher order mutants for reducing the cost of execution of mutants.
They found \emph{each-choice} strategy was the best in this regard.

There have been numerous ideas focused on improving the efficiency of evaluation.
Some of the work in this area involves parallelization of
mutation analysis using MIMD~\cite{offutt1992mutation} and SIMD~\cite{krauser1991high}
machines, in HPC systems~\cite{canizares2016eminent}, and using Hadoop~\cite{saleh2015hadoopmutator}.


\section{Conclusion}
\Mua has been relegated to the sidelines in fuzzing research.
However, we find that two thirds of the key challenges identified in
fuzzing research can potentially be addressed using \mua.

At the same time, we find that \mua still requires solutions to some significant challenges before it can
be used for fuzzing. Of particular imortance is the computational cost.
\Mua is a costly technique even in traditional
settings with static test suites. When used for fuzzing, most of the
traditional optimization techniques can't be used because they assume
static test suites. Further, the fuzzing itself is computationally costly,
with fuzzing campaigns on programs typically recommended to run for at
least 24 hours. Finally, the requirement of statistical confidence
because of non-determinism in fuzzing adds even more computational
requirement for successful use of \mua in fuzzing.

Lack of frameworks that can do both \mua and fuzzing is another problem that
discourages use of \mua in fuzzing. Finally, lack of awareness of \mua
may also be a factor in cybersecurity researchers ignoring \mua for
\fuzzer evaluation.

In this paper, we document the ways in which \mua can potentially help
fuzzing research, or help the fuzzing practitioner, identify the challenges ahead
before \mua can be considered a viable alternative to bug benchmarks and
coverage for \fuzzer evaluation, and propose a few possible mitigation
strategies.



%
\bibliographystyle{IEEEtran}
\bibliography{mutation22-fuzzing}

\end{document}